\documentclass [12pt]{article}
\usepackage{graphicx,amssymb,amsfonts,latexsym,amsmath,amsthm,times}

\usepackage{epsfig}
\usepackage{fancyhdr}

\usepackage[english]{babel}
\usepackage[backend=bibtex8,
            bibstyle=ieee,
            sorting=unsrt,
            citestyle=numeric-comp,
            firstinits=true,
            doi=true,
            isbn=false,
            sorting=none,
            url=false]{biblatex}
\usepackage[breaklinks=true]{hyperref}
\numberwithin{equation}{section}

\newcommand{\pt}{\frac{\partial \rho}{\partial t}}
\newcommand{\rld}[1]{\mathcal{D}_t^{#1}}
\newcommand{\pdx}[1]{\frac{\partial {#1}}{\partial x}}
\newcommand{\pdxs}[1]{\frac{\partial^2 {#1}}{\partial x^2}}

\newtheorem{theorem}{Theorem}
\newtheorem{acknowledgement}[theorem]{Acknowledgement}

\bibliography{ref.bib}
\begin{document}



\vspace*{2cm} \normalsize \centerline{\Large \bf A non-linear subdiffusion model for a cell-cell adhesion in chemotaxis}

\vspace*{1cm}

\centerline{\bf Akram Al-Sabbagh$^{a,b}$\footnote{Corresponding
author. E-mail: akram.al-sabbagh@postgrad.manchester.ac.uk. This work was done as a part of the author's PhD study at the University of Manchester.}}
\vspace*{0.5cm}

\centerline{$^{a}$School of Mathematics, The university of Manchester, Manchester, Oxford Road, M13 9PL, UK}
\centerline {$^{b}$Dept. of Mathematics and Computer Applications, Al-Nahrrain University, 64055 Baghdad, Iraq}


\vspace*{1cm}

\noindent {\bf Abstract.}
The purpose of this work is to propose a non-Markovian and nonlinear model of subdiffusive transport that involves adhesion affects the cells escape rates form position $x$, with chemotaxis. This leads the escape rates to be dependent on the particles density at the neighbours as well as the chemotactic gradient. We systematically derive subdiffusive fractional master equation, then we consider the diffusive limit of the fractional master equation. We finally solve the resulted fractional subdiffusive master equation stationery and analyse the role of adhesion in the resulted stationary density.


\vspace*{1cm}
\setcounter{equation}{0}
\section{Introduction}
In recent years, a stochastic model has become one important tool to deal with the aspect of cell (or organism) migration and adhesion~\cite{turner_discrete_2004,armstrong_continuum_2006,anguige_one-dimensional_2008,simpson_migration_2010,simpson_model_2010,johnston_mean-field_2012,thompson_modelling_2012}. Cell-cell adhesion is a fundamental phenomenon in regard to the subject of cells binding to each other using their surface proteins that are known as cell adhesion molecules~\cite{armstrong_continuum_2006}. Cell migration and adhesion play a key role in many biological phenomena, such as biomolecules~\cite{kendall_molecular_2004}, granular media~\cite{hinrichsen_physics_2004}, cell development~\cite{l_wolpert_principles_2011} and tissue formation, stability and breakdown~\cite{armstrong_continuum_2006} as cells transport to their targets and, by adhesion, they aggregate to from different types of tissues that could be controlled by varying the level of expression of cell adhesion molecules~\cite{foty_differential_2005}.

In general, cell migration can be modelled using two techniques~\cite{erban_individual_2004}, the first is the stochastic (individual–based) model that involves randomness, which is how biological systems behave prevalently, while the other form is deterministic (population-based); this model involves systems of partial differential equations which make it easier to deal with and also more useful for some biological systems~\cite{thompson_modelling_2012}. However, those two forms can be married by understanding the mechanism of cell adhesion that uses the individual-level properties that rules many biological behaviours~\cite{anguige_one-dimensional_2008}.

Cell adhesion models, on the other hand, can be divided into three forms; the discrete approach, which has been widely used to describe many cell migration applications, such as the migration of glioma cell~\cite{deroulers_modeling_2009,khain_collective_2011}, cancer cell invasion~\cite{turner_intercellular_2002,simpson_migration_2010,khain_collective_2011}, wound healing~\cite{anguige_one-dimensional_2008} and many other~\cite{deroulers_modeling_2009,simpson_migration_2010,fernando_nonlinear_2010,penington_building_2011,johnston_mean-field_2012}. The second form of cell adhesion modelling is the continuum model, but there are not many attempts on that subject in the literature~\cite{turner_discrete_2004,armstrong_continuum_2006,anguige_one-dimensional_2008}; some of them were derived by taking the continuous limit of the discrete model. The last type of adhesion modelling is the hybrid model in which both the discrete and the continuous model is combined, and that is even less common~\cite{anderson_hybrid_2005,simpson_model_2010,thompson_modelling_2012}.

The subject of cell movement related to the mean field density has been reported by many authors~\cite{deroulers_modeling_2009,simpson_migration_2010,fernando_nonlinear_2010,penington_building_2011}. It shows the physical behaviour of non-linear diffusive discrete models. However there are few recent attempts in order to derive a CTRW model with nonlinear reaction~\cite{mendez_reaction-transport_2010,angstmann_continuous_2013}. Since the CTRW model has become one popular tool to deal with anomalous diffusion, Fedotov and his collaborators have derived CTRW models for anomalous diffusion process with density dependence jumping rates~\cite{fedotov_nonlinear_2013,straka_transport_2015,falconer_nonlinear_2015}.

The aim in this work is to complete our work in~\cite{falconer_nonlinear_2015} by taking into account of nonlinear dependence on adhesion with anomalous subdiffusion transition in chemotaxis. Therefore, we attempt to derive a non-Markovian random walk model with subdiffusive transport depending on the mean field density together with adhesive effects.
In~\cite{falconer_nonlinear_2015}, we derived the generalised non-linear fractional equation in a diffusion limit as
\begin{align}\label{nonlinear FFPE old}
  \pt &=\pdx{} \left[\frac{l^2}{g(x)} \frac{dS(x)}{dx} e^{-\Phi(x,t)} \rld{1-\nu(x)} \rho(x,t) e^{\Phi(x,t)}-l^2 \kappa \frac{\partial \rho}{\partial x}\alpha(\rho) \right] \notag \\
  &+ \pdxs{} \left[\frac{l^2}{2 g(x)}e^{-\Phi(x,t)} \rld{1-\nu(x)} \rho(x,t) e^{\Phi(x,t)} +\alpha(\rho) \rho(x,t) \right],
\end{align}
Where doth diffusion and transport are regarded to chemotaxis and the mean field density. In this work, however, we assume the adhesion affects the jumping rate to the right and left $\lambda$ and $\mu$. Section~\ref{Non-Markovian nonlinear model sec} presents the derivation of a non-Markovian nonlinear model with density dependence jumping rate, whereas in section~\ref{Diffusion limit and FFPE sec} we derive the generalized master equation in diffusion limit, and in section~\ref{Aggregation and stationary density} we describe the aggregation phenomena in anomalous subdiffusion.


\vspace*{0.5cm}
\setcounter{equation}{0}
\section{Non-Markovian nonlinear model}\label{Non-Markovian nonlinear model sec}

In this section we use the same technique we used in~\cite{falconer_nonlinear_2015} in order to derive the generalized non-linear master equation. We assume a particle moves randomly on a one-dimensional lattice with waiting $T_x$ time between two jumps. The jump direction would be detected by the minimum of two independent random times $T_x^\lambda$ (waiting time preceding jump to the right) and $T_x^\mu$ (waiting time preceding jump to the left). Therefore, the particle jumps to the right if $T_x^\lambda<T_x^\mu$ and jumps to the left otherwise. Then the particle's actual waiting time is
\begin{equation}\label{waiting time def}
T_x=\min\{T_x^\lambda,T_x^\mu \}.
\end{equation}
Where $T_x^\lambda$ and $T_x^\mu$ are distributed as the waiting time PDFs $\psi_\lambda(x,\tau)$ and $\psi_\mu(x,\tau)$ respectively, where $\tau$ is the particle's residence time at position $x$. The waiting time PDFs $\psi_\lambda(x,\tau)$ and $\psi_\mu(x,\tau)$ are defined as the limit
\begin{equation}\label{waiting time pdf limts}
\psi_\lambda(x,\tau)=\lim_{\Delta \tau \rightarrow 0}\frac{\Pr\{\tau<T_x^\lambda<\tau+\Delta t \}}{\Delta t},\ \ \ \ \ \psi_\mu(x,\tau)=\lim_{\Delta \tau \rightarrow 0}\frac{\Pr\{\tau<T_x^\mu<\tau+\Delta t \}}{\Delta t}.
\end{equation}
Accordingly, the survival function PDFs of jumping to the right and left are defined relative to the waiting time PDFs as
\begin{equation}\label{survival pdf's limit}
\Psi_\lambda(x,\tau)=\int_{\tau}^{\infty} \psi_\lambda(x,s)ds,\ \ \ \ \ \Psi_\mu(x,\tau)=\int_{\tau}^{\infty} \psi_\mu(x,s)ds.
\end{equation}
At this stage, we will introduce the modified escape rates of particles to the right $\lambda_\alpha$ and to the left $\mu_\alpha$, where these escape rates are now depend on the density of particles at the neighbour as well as the particle position $x$ and its residence time $\tau$. They are defined as
\begin{equation}\label{escape rates}
\lambda_\alpha=\lambda(x,\tau)+\alpha_\lambda(\rho(x-l,t)),\ \ \ \ \ \mu_\alpha=\mu(x,\tau)+\alpha_\mu(\rho(x+l,t)).
\end{equation}
Here, $\lambda(x,\tau)$ and $\mu(x,\tau)$ represent the linear escape rates with no mean field density dependence and are defined as
\begin{equation}\label{linear escape rates}
\lambda(x,\tau)=\lim_{\Delta \tau \rightarrow 0} \frac{\Pr\{T_x^\lambda<\tau+\Delta t|T_x^\lambda>\tau \}}{\Delta t},\ \ \ \ \ \mu(x,\tau)=\lim_{\Delta \tau \rightarrow 0} \frac{\Pr\{T_x^\mu<\tau +\Delta t|T_x^\mu>\tau \}}{\Delta t},
\end{equation}
And $\alpha_\lambda(\rho(x-l,t))$ and $\alpha_\mu(\rho(x+l,t))$ are the addition nonlinear escape rates with adhesion effects. The nonlinear term in escape rates are independent of the anomalous trapping with probability of escaping equal to $\alpha(\rho)\Delta t$ in a time interval of $\Delta t$, and
\begin{equation}\label{total nonlinear escape rates}
\alpha(\rho(x,t))=\alpha_\lambda(\rho(x-l,t))+\alpha_\mu(\rho(x+l,t)).
\end{equation}
This flexible form of escape rates leads us to generalize several nonlinear effects, for instance by the following choice of escape rates,
\begin{equation}\label{volume filling}
\lambda_\alpha=\lambda(x,\tau)+\alpha_\lambda(\rho(x+l,t)),\ \ \ \ \ \mu_\alpha=\mu(x,\tau)+\alpha_\mu(\rho(x-l,t)),
\end{equation}
the volume filling effects can be obtained, where particles that have a non-zero volume prevent other particles from diffusing through the occupied area~\cite{hillen_global_2001,hillen_users_2008}.
Also, those escape rates could lead to the model of the local gradient of density~\cite{hillen_users_2008} by choosing~\cite{falconer_nonlinear_2015}
\begin{equation}\label{mean density escape rates}
\lambda_\alpha=\lambda(x,\tau)+\kappa (\alpha_\lambda(\rho(x+l,t))-\alpha_\lambda(\rho(x,t))),\ \ \ \ \ \mu_\alpha=\mu(x,\tau)+\kappa(\alpha_\mu(\rho(x,t))-\alpha_\mu(\rho(x-l,t))).
\end{equation}
The escape rates $\lambda$ and $\mu$ could also be written in terms of both the waiting time PDF and the survival function, by the use of Bayes' theorem, as
\begin{equation}\label{escape rates waiting time/survival}
\lambda(x,\tau)=\frac{\psi_\lambda(x,\tau)}{\Psi(x,\tau)},\ \ \ \ \ \mu(x,\tau)=\frac{\psi_\mu(x,\tau)}{\Psi(x,\tau)},
\end{equation}
Where $\Psi(x,\tau)$ represents the total survival PDF, which is basically the product of the two survival functions $\Psi_\lambda$ and $\Psi_\mu$, and is also defined as
\begin{equation}\label{survival pdf}
\Psi(x,\tau)=e^{-\int_0^\tau (\lambda(x,s)+\mu(x,s))ds}.
\end{equation}
The total waiting time PDF $\psi(x,\tau)$, in addition, is defined to be the summation of the waiting time PDF of jumping to the right $\psi_\lambda(x,\tau)$ and to the left $\psi_\mu(x,\tau)$, that are defined as
\begin{equation}\label{waiting time pdfs}
\psi_\lambda(x,\tau)=\frac{-\partial \Psi_\mu(x,\tau)}{\partial \tau} \Psi_\lambda(x,\tau),\ \ \ \ \ \psi_\mu(x,\tau)=\frac{-\partial \Psi_\lambda(x,\tau)}{\partial \tau} \Psi_\mu(x,\tau).
\end{equation}
In order to derive the generalized master equation for the non-Markovian model, we follow the procedure of adding an auxiliary variable $\tau$ to generate a structured density $\xi(x,t,\tau)$ that represent the particle density at position $x$ at time $t$ being trapped for time $\tau$~\cite{mendez_reaction-transport_2010,vlad_systematic_2002,yadav_kinetic_2006,falconer_nonlinear_2015}. This structured density should obey the balance equation
\begin{align}\label{structured balance equation}
\frac{\partial \xi}{\partial t}+\frac{\partial \xi}{\partial \tau} &= -(\lambda(x,\tau)+\alpha_\lambda(\rho(x-l,t))\xi(x,t,\tau) \notag \\
&-(\mu(x,\tau)+\alpha_\mu(\rho(x+l,t))\xi(x,t,\tau).
\end{align}
The target now is to derive the general form of the unstructured density $\rho(x,t)$, where
\begin{equation}\label{unstructured= int structured}
\rho(x,t)=\int_0^t \xi(x,t,\tau) d\tau.
\end{equation}
This approach has been found to be one useful technique to deal with non-Markovian random walk~\cite{fedotov_subdiffusion_2011,fedotov_subdiffusive_2012,fedotov_non-homogeneous_2013,vlad_systematic_2002,yadav_kinetic_2006}, especially for the nonlinear generalizations~\cite{fedotov_nonlinear_2013,fedotov_nonlinear_2014,fedotov_subdiffusion_2015,straka_transport_2015,falconer_nonlinear_2015}.
The balance equation~\eqref{structured balance equation} needs to satisfy the initial condition for $t=0$ that is given by
\begin{equation}\label{structured initial condition}
\xi(x,0,\tau)=\rho_0(x)\delta(\tau),
\end{equation}
Where $\rho_0(x)$ is the initial density of particles and $\delta$ is the Dirac delta function. Also, the boundary condition where the residence time $\tau=0$ is given by
\begin{align}\label{boundary condition}
\xi(x,t,0) &=\int_0^t [\lambda(x-l,\tau)+\alpha_\lambda(\rho(x-2l,t)) ] \xi(x-l,t,\tau) d\tau \notag \\
&+\int_0^t [\mu(x+l,\tau)+\alpha_\mu(\rho(x+2l,t)) ] \xi(x+l,t,\tau) d\tau.
\end{align}
Denote the integral arrival rate as $j(x,t)$, which represents the rate of particles that arrive to position $x$ at exactly time $t$. This rate is equivalent to the boundary condition above~\eqref{boundary condition}, that is
\begin{equation}\label{intigral arrival rate def}
j(x,t)=\xi(x,t,0).
\end{equation}
We introduce the integral escape rates to the right $i_\lambda$ and to the left $i_\mu$ as
\begin{align}\label{integral escape rates def}
i_\lambda(x,t)&=\int_0^t \lambda_\alpha \xi(x,t,\tau) d\tau, \notag \\
i_\mu(x,t)&=\int_0^t \mu_\alpha \xi(x,t,\tau) d\tau.
\end{align}
Then the boundary condition~\eqref{boundary condition} can now be presented as
\begin{equation}
j(x,t)=i_\lambda(x-l,t)+i_\mu(x+l,t).
\end{equation}
In order to get the final form for the generalised master equation for the unstructured density, we first solve the balance equation~\eqref{structured balance equation} using the method of characteristics and get
\begin{equation}\label{char solution}
\xi(x,t,\tau)=\xi(x,t-\tau,0) e^{-\int_0^\tau (\lambda(x,s)+\mu(x,s))ds} e^{-\int_{t-\tau}^t \alpha(\rho)ds},
\end{equation}
or in other words,
\begin{equation}\label{characteristics solution}
\xi(x,t,\tau)=j(x,t-\tau)\Psi(x,\tau) e^{-\int_{t-\tau}^t \alpha(\rho)ds},
\end{equation}
where $\alpha(\rho)=\alpha_\lambda(\rho(x-l,t))+\alpha_\mu(\rho(x+l,t))$. Inserting the characteristics solution~\eqref{characteristics solution} into~\eqref{unstructured= int structured}, and by the use of the initial condition~\eqref{structured initial condition}, one can get
\begin{equation}\label{unstructured density solution}
\rho(x,t)=\int_0^t \Psi(x,\tau) j(x,t-\tau)e^{-\int_{t-\tau}^t \alpha(\rho)ds} d\tau + \rho_0(x) \Psi(x,t) e^{-\int_0^t \alpha(\rho)ds}.
\end{equation}
Hence, the integral escape rates to the right and left in equation~\eqref{integral escape rates def} can be represented  using the solution of~\eqref{characteristics solution} in the form
\begin{align}\label{integral escape rates solution}
i_\lambda(x,t) &=\int_0^t \psi_\lambda(x,\tau) j(x,t-\tau) e^{-\int_{t-\tau}^t \alpha(\rho)ds} d\tau \notag \\
 &+\rho_0(x) \psi_\lambda(x,t) e^{-\int_0^t \alpha(\rho)ds} +\alpha_\lambda(\rho(x-l,t)) \rho(x,t), \notag \\
i_\mu(x,t) &=\int_0^t \psi_\mu(x,\tau) j(x,t-\tau) e^{-\int_{t-\tau}^t \alpha(\rho)ds} d\tau \notag \\
 &+\rho_0(x) \psi_\mu(x,t) e^{-\int_0^t \alpha(\rho)ds} +\alpha_\mu(\rho(x+l,t)) \rho(x,t).
\end{align}
Finally, to eliminate the expression of the arrival rate $j(x,t)$ from the above form, we apply the Laplace transform on~\eqref{integral escape rates solution} and then invert the transport to obtain
\begin{align}\label{integral escape rates with kernel}
i_\lambda(x,t) &=\int_0^t K_\lambda(x,t-\tau) \rho(x,\tau) e^{-\int_0^t \alpha(\rho)ds}d\tau +\alpha_\lambda(\rho(x-l,t)) \rho(x,t), \notag \\
i_\mu(x,t) &=\int_0^t K_\mu(x,t-\tau) \rho(x,\tau) e^{-\int_0^t \alpha(\rho)ds}d\tau +\alpha_\mu(\rho(x+l,t)) \rho(x,t),
\end{align}
where $K_\lambda$ and $K_\mu$ are the memory kernels  that are defined in Laplace space as
\begin{equation}\label{memory kernels}
\hat{K}_\lambda(x,s)=\frac{\hat{\psi}_\lambda(x,s)}{\hat{\Psi}(x,s)},\ \ \ \ \ \hat{K}_\mu(x,s)=\frac{\hat{\psi}_\mu(x,s)}{\hat{\Psi}(x,s)}.
\end{equation}
In order to get the final approach of the master equation for the unstructured density, we differentiate~\eqref{unstructured= int structured} with respect to $t$, and use the balance equation of the structured density~\eqref{structured balance equation} to get
\[\pt=j(x,t)-i_\lambda(x,t)-i_\mu(x,t)\]
or
\begin{equation}\label{unstructured master equation def}
\pt=i_\lambda(x-l,t)+i_\mu(x+l,t)-i_\lambda(x,t)-i_\mu(x,t).
\end{equation}
This equation is described as the balance of particles that arrive to and leave from the state $x$ at time $t$. Therefore, the final expression of the generalised master equation for the unstructured density can be formulated as
\begin{align}\label{unstructured generalised master equation}
\pt&=\int_0^t K_\lambda(x-l,t-\tau) \rho(x-l,\tau) e^{-\int_0^t \alpha(\rho(x-l,t))ds} d\tau +\alpha_\lambda(\rho(x-2l,t)) \rho(x-l,t) \notag \\
 &+ \int_0^t K_\mu(x+l,t-\tau) \rho(x+l,\tau) e^{-\int_0^t \alpha(\rho(x+l,t))ds} d\tau +\alpha_\mu(\rho(x+2l,t)) \rho(x+l,t) \notag \\
 &-\int_0^t K(x,t-\tau) \rho(x,\tau) e^{-\int_0^t \alpha(\rho)ds} d\tau -\alpha(\rho) \rho(x,t),
\end{align}
where $K(x,t)=K_\lambda(x,t)+K_\mu(x,t)$ is the combined memory kernel and $\alpha(\rho)=\alpha_\lambda(\rho(x-l,t))+\alpha_\mu(\rho(x+l,t))$.


\vspace*{0.5cm}
\setcounter{equation}{0}
\section{Anomalous subdiffusion model}\label{Anomalous subdiffusion model sec}

In this section we deal with a special case of the generalised master equation~\eqref{unstructured generalised master equation} and derive the fractional master equation for our model. Assuming that the escape rates are inversely proportional to the residence time $\tau$ leads to a time fractional memory kernel, that is
\begin{equation}\label{anomalous escape rates}
\lambda(x,\tau)=\frac{\nu_\lambda(x)}{\tau_0(x)+\tau},\ \ \ \ \ \mu(x,\tau)=\frac{\nu_\mu(x)}{\tau_0(x)+\tau}.
\end{equation}
Recalling the definition~\eqref{escape rates waiting time/survival}, the survival function now have a power-law dependence~\cite{falconer_nonlinear_2015} defined in
\begin{equation}\label{anomalous survival PDFs}
\Psi_\lambda(x,\tau)=\left[\frac{\tau_0(x)}{\tau_0(x)+\tau}\right]^{\nu_\lambda(x)},\ \ \ \ \ \Psi_\mu(x,\tau)=\left[\frac{\tau_0(x)}{\tau_0(x)+\tau}\right]^{\nu_\mu(x)},
\end{equation}
and this leads to the total survival PDF
\begin{equation}\label{anomalous total survival pdf}
\Psi(x,\tau)=\left[\frac{\tau_0(x)}{\tau_0(x)+\tau}\right]^{\nu(x)},
\end{equation}
where $\nu_\lambda(x)$ and $\nu_\mu(x)$ represent the anomalous exponent of jumping to the right and left respectively and $\nu(x)$ is the total anomalous exponent, all of which only depend on the spatial variable $x$. On the other hand, the total waiting time PDF $\psi(x,\tau)$ has a Pareto density in the anomalous case,
\begin{equation}\label{anomalous total waiting time pdf}
\psi(x,\tau)=\frac{\nu(x)\tau_0(x)^{\nu(x)}}{(\tau_0(x)+\tau)^{1-\nu(x)}}.
\end{equation}
At this stage, let us introduce the jump probabilities that are independent of the residence time $\tau$ as a ratio of the escape rates $\lambda$ or $\mu$ to the sum of them, $\lambda+\mu$, as follows:
\begin{equation}\label{jump probabilities}
p_\lambda(x)=\frac{\nu_\lambda(x)}{\nu(x)},\ \ \ \ \ p_\mu(x)=\frac{\nu_\mu(x)}{\nu(x)}.
\end{equation}
This leads, by the use of the Tauberian theorem, to the new Laplace form of the memory kernels
\begin{equation}\label{anomalous memory kernels}
\hat{K}_\lambda(x,s) \simeq \frac{p_\lambda(x) s^{1-\nu(x)}} {g(x)},\ \ \ \ \ \hat{K}_\mu(x,s) \simeq \frac{p_\mu(x) s^{1-\nu(x)}} {g(x)},
\end{equation}
and since $p_\lambda(x)+p_\mu(x)=1$, then the total memory kernel is
\begin{equation}\label{anomalous total memory kernel}
\hat{K}(x,s)=\frac{s^{1-\nu(x)}}{g(x)}.
\end{equation}
The integral escape rates in~\eqref{integral escape rates with kernel} can now be redefined in terms of the memory kernels with anomalous exponents as
\begin{align}\label{anomalous integral escape rates}
i_\lambda(x,t) &=a(x) e^{-\int_0^t \alpha(\rho)dt} \rld{1-\nu(x)} \left[\rho(x,t) e^{\int_0^t \alpha(\rho)dt} \right] +\alpha_\lambda(\rho(x-l,t)) \rho(x,t), \notag \\
i_\mu(x,t) &=b(x) e^{-\int_0^t \alpha(\rho)dt} \rld{1-\nu(x)} \left[\rho(x,t) e^{\int_0^t \alpha(\rho)dt} \right] +\alpha_\mu(\rho(x+l,t)) \rho(x,t),
\end{align}
therefore, the fractional master equation can be presented as
\begin{align}\label{fractional master equation}
\pt &=a(x-l) e^{-\int_0^t \alpha(\rho(x-l,t))dt} \rld{1-\nu(x-l)} \left[\rho(x-l,t) e^{\int_0^t \alpha(\rho(x-l,t))dt} \right] +\alpha_\lambda(\rho(x-2l,t)) \rho(x-l,t) \notag \\
 &+b(x+l) e^{-\int_0^t \alpha(\rho(x+l,t))dt} \rld{1-\nu(x+l)} \left[\rho(x+l,t) e^{\int_0^t \alpha(\rho(x+l,t))dt} \right] +\alpha_\mu(\rho(x+2l,t)) \rho(x+l,t) \notag \\
 &-[a(x)+b(x)] e^{-\int_0^t \alpha(\rho)dt} \rld{1-\nu(x)} \left[\rho(x,t) e^{\int_0^t \alpha(\rho)dt} \right] -\alpha(\rho) \rho(x,t).
\end{align}


\vspace*{0.5cm}
\setcounter{equation}{0}
\section{Diffusion limit and FFPE}\label{Diffusion limit and FFPE sec}

In this section we attempt to derive the FFPE in a diffusive limit. For the limit of $l\rightarrow 0$, let us define the flux of particles from $x\rightarrow x+l$, denoted by $J_\lambda$, as
\begin{equation}\label{right flux}
J_\lambda= i_\lambda(x,t)l-i_\mu(x+l,t)l,
\end{equation}
and also, the flux of particles from $x-l\rightarrow x$, denoted by $J_\mu$, as
\begin{equation}\label{left flux}
J_\mu=i_\lambda(x-l,t)l-i_\mu(x,t)l.
\end{equation}
Recall the formula~\eqref{unstructured master equation def} of the master equation. By the use of the flux definitions~\eqref{right flux} and~\eqref{left flux}, one can have
\begin{equation}
\pt=\frac{-J_\lambda +J_mu}{l},
\end{equation}
and for the limit of $l\rightarrow 0$, we get
\begin{equation}\label{master in flux}
\pt=-\pdx{J(x,t)},
\end{equation}
where the operator $J(x,t)$ is defined as~\cite{fedotov_nonlinear_2013}
\begin{equation}\label{total flux}
J(x,t)=-\frac{l^2}{2}\pdx{i} -\frac{l^2\ (i_\lambda-i_\mu)}{1-\rho} \pdx{\rho}+o(a^2).
\end{equation}

In what follows, we consider the probability of jumping to the right $p_\lambda$ and to the left $p_\mu$ both depend on the chemotactic substance $S(x)$, and are defined as
\begin{equation}\label{linjumpprob}
p_\lambda(x)=Ae^{-\beta (S(x+l)-S(x))},\ \ \ \ \ p_\mu(x)=Ae^{-\beta (S(x-l)-S(x))},
\end{equation}
where $A$ can be calculated from $p_\lambda(x)+p_\mu(x)=1$, and the difference between the two probabilities is approximated by~\cite{fedotov_non-homogeneous_2013,falconer_nonlinear_2015}
\begin{equation}\label{linear jump prob with chemotaxis}
p_\lambda(x)-p_\mu(x)=-l\beta\frac{dS(x)}{dx}+o(l).
\end{equation}
Also, we can organise the difference between $\alpha_\lambda$ and $\alpha_\mu$ in an analogous manner. If we suggest
\[p_{\alpha,\lambda}(x,t)=\frac{\alpha_\lambda(\rho(x-l,t))}{\alpha(\rho(x,t))},\ \ \ \ \ p_{\alpha,\mu}(x,t)=\frac{\alpha_\mu(\rho(x+l,t))}{\alpha(\rho(x,t))}, \]
and choose $\alpha_\lambda$ and $\alpha_\mu$ such as
\begin{equation}
p_{\alpha,\lambda}(x,t)=Be^{\kappa(\rho(x+l,t)-\rho(x,t))},\ \ \ \ \ p_{\alpha,\mu}(x,t)=Be^{\kappa(\rho(x-l,t)-\rho(x,t))},
\label{nonlinjumpprob}
\end{equation}
where $B$ satisfies $p_{\alpha,\lambda}(x,t)+p_{\alpha,\mu}(x,t)=1$, the approximation of the difference can be illustrated by
\begin{equation}\label{nonlinear jump prob}
p_{\alpha,\lambda}(x,t)-p_{\alpha,\mu}(x,t)=-l\kappa\frac{\partial \rho}{\partial x}+o(l).
\end{equation}
Thus, using the expression~\eqref{master in flux}, by the use of~\eqref{total flux},~\eqref{linear jump prob with chemotaxis} and~\eqref{nonlinear jump prob}, the fractional master equation~\eqref{fractional master equation} can now be written in the form
\begin{align}\label{FFPE with adhesion}
\pt &= l^2 \pdx{} \frac{1}{1-\rho(x,t)} \pdx{\rho} \left[\frac{\beta}{g(x)} \frac{dS(x)}{dx} e^{-\Phi(x,t)} \mathcal{D}_t^{1-\nu(x)}[\rho(x,t) e^{\Phi(x,t)}]+\kappa \frac{\partial \rho}{\partial x} \alpha(\rho(x,t)) \right] \notag \\
&+\frac{l^2}{2} \frac{\partial^2}{\partial x^2} \left[\frac{e^{-\Phi(x,t)}}{g(x)} \mathcal{D}_t^{1-\nu(x)}[\rho(x,t) e^{\Phi(x,t)}] +\alpha(\rho(x,t))\rho(x,t) \right],
\end{align}
where $\Phi(x,t)=\int_0^t \alpha(\rho(x,t))dt$ and $\alpha(\rho(x,t))=\alpha_\lambda(\rho(x-l,t))+\alpha_\mu(\rho(x+l,t))$.


\vspace*{0.5cm}
\setcounter{equation}{0}
\section{Aggregation and stationary density}\label{Aggregation and stationary density}

The attempt in this section is to find the stationary form of the master equation~\eqref{FFPE with adhesion}, where the non-linear escape rates $\alpha_\lambda(\rho)$ and $\alpha_\mu(\rho)$ depend on the particles density at the point of next jump position. Therefore, by the definition of the stationary density, in the limit of $s\rightarrow 0$
\[\rho_{st}(x)=\lim_{s\rightarrow 0} s \hat{\rho}(x,s), \]
and the stationary integral escape rates are
\[\hat{i}_\lambda(x)=\lim{s\rightarrow 0} s \hat{i}_\lambda(x,s),\ \ \ \ \ \hat{i}_\mu(x)=\lim{s\rightarrow 0} s \hat{i}_\mu(x,s). \]

\begin{figure}[htp]
	\begin{center}
		\includegraphics[width=3in]{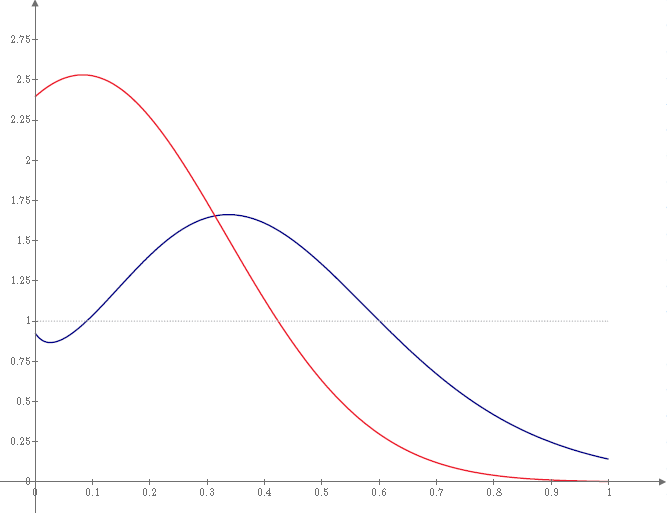}
		\caption{Stationary density of particles $\rho{st}$ in quadratic chemotaxis gradient $S(x)=5x$, on domain of $[0,1]$, calculated with anomalous substance $\nu(x)=\nu_0 e^{-kx}$, $\nu_0=0.9$ and $k=2.19$. The blue line represent the density with no adhesion effects $\lambda_\aleph=\mu_\alpha=2\times10^{-4}(1-\chi \rho_{st})$. The red line is the density where escape rates depend on the density of particles on the neighbours $\lambda_\aleph=2\times10^{-4}(1-\chi \rho_{st,-1})$, and $\mu_\aleph=2\times10^{-4}(1-\chi \rho_{st,+1})$, where $\chi=0.1$.}
		\label{fig1}
	\end{center}
\end{figure}

Assuming that $e^{-\int_0^t \alpha(\rho)dt}$ can be approximated by $e^{-\alpha(\rho_{st}(x))t}$, the stationary integral escape rates can be written in the form~\cite{falconer_nonlinear_2015}
\begin{equation}\label{stationary integral escape rates}
i_{\lambda,st}(x)=\lambda_\nu (x) \rho_{st}(x),\ \ \ \ \ i_{\mu,st}(x)=\mu_\nu (x) \rho_{st}(x),
\end{equation}
where
\begin{align}\label{(anomalous)}
\lambda_\nu(x) &=a(x) \alpha(\rho_{st}(x))^{1-\nu(x)} +\alpha_\lambda(\rho_{st}(x-l)), \notag \\
\mu_\nu(x) &=b(x) \alpha(\rho_{st}(x))^{1-\nu(x)} +\alpha_\mu(\rho_{st}(x+l)).
\end{align}
Then the fractional master equation~\eqref{FFPE with adhesion} can be rewritten in terms of the stationary density, with vanishing time derivative $\pt\rightarrow 0$, as
\[0= -\pdx{}\left[\frac{-l^2}{2} \pdx{i_{st}} + \frac{l^2}{1-\rho_{st}} (i_{\lambda,st}-i_{\mu,st})\pdx{\rho} \right],  \]
and hence
\begin{align}
0 &= \pdxs{} \frac{l^2 \alpha(\rho_{st}(x))^{1-\nu(x)}}{2} \left[\frac{1}{g(x)}+\alpha(\rho_{st}) \right]\rho_{st}  \notag \\
&- \pdx{} \frac{l^2}{1-\rho_{st}} \pdx{\rho_{st}} \left[\frac{p_\lambda(x)-p_\mu(x)}{g(x)} \alpha(\rho_{st})^{1-\nu(x)} +\kappa \pdx{\rho_{st}} \right] \rho_{st}.
\end{align}
Substituting the values of~\eqref{linear jump prob with chemotaxis} and~\eqref{nonlinear jump prob}, we get the final form
\begin{equation}\label{stationary FFPE}
\pdx{}D_\nu(x,\rho_{st}) \rho_{st}(x)=v(x,\rho_{st}) \rho_{st} \pdx{\rho_{st}},
\end{equation}
where $D_\nu(x,\rho_{st})$ is the diffusion function, and $v(x,\rho_{st})$ is the velocity function, and are defined as
\begin{equation}\label{diffusion function}
D_\nu(x,\rho_{st})=\frac{l^2 \alpha(\rho_{st}(x))^{1-\nu(x)}}{2} \left[\frac{1}{g(x)}+\alpha(\rho_{st}) \right],
\end{equation}
and
\begin{equation}
v(x,\rho_{st})=\frac{l^2}{1-\rho_{st}} \left[\frac{\beta}{g(x)} \frac{d S(x)}{dx} \alpha(\rho_{st})^{1-\nu(x)} +\kappa \pdx{\rho_{st}} \right].
\end{equation}
The final fractional master equation~\eqref{stationary FFPE} can be numerically approximated with some specific assumptions as it is complicated and not easy to solve. Figure~\eqref{fig1} represents the simulated stationary density $\rho_{st}$ distributed over the interval $[0,1]$; it shows that adhesion effects on particles seemed to slow them down and force them to aggregate on the boundary.

\vspace*{0.5cm}

\section*{Conclusions}
The main target in this work has been to implement adhesion effect, involving nonlinear dependence, into fractional subdiffusion transition model. We presented a non-Markovian and non-linear random walk model, where the escape rates inversely proportional to the residence time $\tau$, and they also depend on the density of particles at the neighbours. We derived the generalised master equation for this model, it included the exponential factor that involves the non-linear escape rate dependence with adhesion effects. In the subdiffusive case, the resulted master equation involved to a non-trivial combination of the non-linear exponential factor together with the Riemann-Liouville fractional derivative. This combination performs as a tempering to the anomalous trapping process.
In long time limit, we could evaluate the fractional master equation in terms of stationary density. The stationary solution with non-linear dependence of $\alpha$ upon to $\rho$ led to a nonlinear advection and diffusion, which are both dependent on nonlinear escape rates with adhesion effects. Finally, we presented a numerical simulation for the stationary solution in figure~\eqref{fig1}, that showed that adhesion effects seemed to slow particles down and force them to aggregate. This Model can also be applied on transport systems with non-standard diffusion, where memory kernels causes a dependence between reactions and transport, like the case of propagating front~\cite{yadav_propagating_2007}.


\begin{acknowledgement}
 The author is very grateful to Prof. Sergei Fedotov for his great help and support.
\end{acknowledgement}


\printbibliography

\end{document}